\documentclass[aps,pra,reprint,groupedaddress,footinbib,showpacs]{revtex4-1}

\usepackage{graphicx}
\usepackage{upgreek} 
\usepackage{amsmath} 

\usepackage{pifont}
\newcounter{mycounter}
\renewcommand*\textcircled[1]{\protect\setcounter{mycounter}{#1}\protect\addtocounter{mycounter}{191}\protect\raisebox{-2pt}{\Large\ding{\themycounter}}}

\newcommand{\ket}[1]{\mbox{\ensuremath{\mid\!\! #1 \rangle}}}
\begin{document}


\title{Motion-induced signal revival in pulsed Rydberg four-wave mixing \\ beyond the frozen gas limit} 


\author{B. Huber}
\affiliation{5. Physikalisches Institut, Universit\"{a}t Stuttgart,
Pfaffenwaldring 57, 70550 Stuttgart, Germany}
\author{A. K\"olle}
\affiliation{5. Physikalisches Institut, Universit\"{a}t Stuttgart,
Pfaffenwaldring 57, 70550 Stuttgart, Germany}
\author{T. Pfau}
\affiliation{5. Physikalisches Institut, Universit\"{a}t Stuttgart,
Pfaffenwaldring 57, 70550 Stuttgart, Germany}


\date{\today}

\begin{abstract}
We present measurements on pulsed four-wave mixing involving a Rydberg state in an atomic vapor cell. 
The excitation to the Rydberg state is conducted by a pulsed two-photon excitation on the nanosecond timescale that is combined with a third CW laser in phase-matched geometry yielding light emission on the same timescale.
An additional signal peak is observed shortly after the pulse 
that can be attributed to a revival of constructive interference between different velocity classes of the radiating atomic dipoles.
Furthermore we investigate the density dependence of the four-wave mixing signal. From the shape of the respective curve 
we are able to confirm
energy and momentum conservation in the photonic part of the system.
\end{abstract}


\maketitle

\section{Introduction}
A physical process involving four light fields can be used to transfer information of a light field to a medium in the form of spatially extended spin waves and to subsequently retrieve it. This effect is widely used for a variety of applications including atomic quantum memories based on read/write sequences \cite{Lukin2000,Julsgaard2004,Matsukevich2004,Eisaman2005,Hosseini2009}, where the fields are temporally separated, or  -- based on four-wave mixing -- the creation of exotic wavelengths \cite{Scheid2009}, wavelength conversion \cite{Radnaev2010} or the production of correlated photon pairs \cite{Chaneliere2006, Willis2011}. 
Additionally, non-linearities of the medium itself can be exploited to impose non-classical features onto the light field created, e.g. squeezed states of light \cite{Slusher1985}. 
The strong interaction between Rydberg states \cite{Tong2004,Singer2004,Vogt2006} can be used to enhance the non-linearities of the medium \cite{Friedler2005, Pritchard2010, Peyronel2012}, on the basis of which a source of single photons has been proposed \cite{Saffman2002, Saffman2010} and realized in an ultracold atomic ensemble \cite{Dudin2012a}.

For Rydberg excitation in thermal vapor, however, a broad distribution of velocity classes is present, which leads to motion-induced decoherence on the nanosecond timescale due to the Doppler effect. While narrow-band coherent Rydberg excitation is possible for the observation of steady state phenomena like EIT \cite{Mohapatra2007} or CW four-wave mixing \cite{Kolle2012}, significantly larger bandwidths are favorable for the study of coherent dynamics. Coherent dynamics in the frozen gas regime, where the excitation bandwidth exceeds the Doppler width, has been demonstrated previously \cite{Huber2011a}.

In this letter, we investigate the dynamics beyond the frozen gas regime, where the atomic velocities still have a significant influence on the time evolution. 
We present measurements of a pulsed four-wave mixing experiment via a Rydberg state in atomic vapor above room temperature. 
The excitation to the Rydberg state is conducted via a two-photon excitation in a pulsed manner within a duration of 
few nanoseconds, while the final state is weakly coupled by a CW laser. We observe light emission on the fourth transition during the time of the excitation. As the excitation bandwidth to the Rydberg state is below the respective Doppler width, no single-atom coherent dynamics like e.g. Rabi flopping \cite{Huber2011a} can be expected from the Doppler ensemble. However, within a certain range of the excitation Rabi frequency we observe two signal peaks, where the second peak occurs after the excitation pulse has passed the atomic sample. This second signal peak can be attributed to a revival of constructive interference between radiating atomic dipoles of different velocity classes that are excited within the thermal ensemble. 
The interplay of the different velocity classes 
will be discussed in detail with the help of a four-level model to describe this non-trivial temporal dynamics in the signal. We find good agreement with the model over a range of different Rabi frequencies. 


Density-dependent measurements are conducted to investigate the re-absorption of the four-wave mixing signal in the atomic medium. We find that the magnitude of the signal can be well described by a simple model based on Lambert-Beer's absorption law. 
The behavior of the signal strength for different optical densities allows to draw conclusions about energy and momentum conservation in the photonic part of the system. 

\section{Experimental situation}

We use a vapor cell filled with rubidium at natural abundance ($72.2\%$ $^{85}$Rb, $27.8\%$ $^{87}$Rb). The thickness of the vapor-filled volume is $\sim\,680\,\mathrm{\upmu m}.$

\begin{figure}
\centering
\includegraphics[width=0.65\columnwidth]{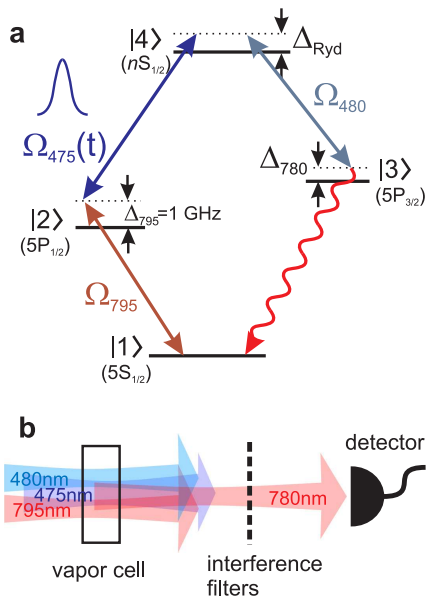}
\caption{(a) Diamond excitation scheme. We use a pulsed two-photon-excitation scheme for addressing the Rydberg state $n\mathrm{S}_{1/2}$ via the D1-line.
A third CW laser couples the Rydberg level to the $5\mathrm{P}_{3/2}$ state, from which the atoms radiate coherently to the ground state completing the four-wave mixing process.
(b) Optical setup. All three beams are overlapped in the vapor cell in co-propagation configuration thereby fulfilling the phase-matching condition. The four-wave mixing signal at $780\,\mathrm{nm}$ is separated from the excitation light by a stack of optical bandpass filters before detection.
\label{fig:schema}}
\end{figure}

The four-wave mixing process is conducted in a diamond excitation scheme (fig.~\ref{fig:schema}a). We address the Rydberg state by an effective two-photon transition via the D1 line. 
The laser at $795\,\mathrm{nm}$ is locked blue detuned by $\Delta_{795}/2\pi = 1\,\mathrm{GHz}$ with respect to the $\,5\mathrm{S}_{1/2},F=3\rightarrow 5\mathrm{P}_{1/2},F'=3$ transition ($^{85}\mathrm{Rb}$).
The adjacent transition to the Rydberg state is coupled by a \mbox{$\sim\! 2\,\mbox{-}\,3\,\mathrm{ns}$} pulse at $475\,\mathrm{nm}$ produced by a seeded dye laser amplifier similar to \cite{Schwettmann2007}. Both lasers provide Rabi frequencies in the range of several hundred MHz. 
The Rydberg state, in turn, is weakly coupled on the way down to the $5\mathrm{P}_{3/2}$ state by a CW laser at $480\,\mathrm{nm}$ ($\Omega_{480}/2\pi \lesssim 10\,\mathrm{MHz}$, depending on the Rydberg state). This laser is locked resonantly to the respective transition with reference to the center of gravity of the $\,5\mathrm{S}_{1/2},F=3$ D2 hyperfine lines. 
In phase-matched configuration this leads to coherent and directed light emission on the D2 line at $780\,\mathrm{nm}$.
All lasers share the same linear polarization.

We fulfill the phase-matching condition spatially with a fully co-propagating alignment of the three laser beams (fig.~\ref{fig:schema}b).
The light is separated from the excitation lasers by a stack of optical bandpass filters and subsequently recorded by a single photon counting module. The time evolution of the signal, therefore, has to be acquired by statistical measurements of the photon delay relative to the laser pulse. Although the signal strength is well above single photon level and thus has to be attenuated accordingly, we have chosen this detection method for its good time resolution and sensitivity. 

\section{Model}

We describe the single-atom dynamics with a four-level model coupled by three different light fields. Using a density matrix approach \cite{Fleischhauer2005}, the time evolution is given by the master equation $\dot{\hat{\rho}}=-\frac{i}{\hbar}[\hat{H},\hat{\rho}]+L(\hat{\rho})$. The corresponding Hamiltonian in the rotating wave approximation reads
\begin{equation}
\hat{H} = \hbar
\begin{pmatrix}
0 												& \frac{1}{2}\Omega_{795}			& 0 													& 0\\
\frac{1}{2}{\Omega_{795}}^* & - \Delta_{795} 						& 0 													& \frac{1}{2}\Omega_{475}(t)	\\
0 												& 0													 	& - \Delta_{780}							& \frac{1}{2}\Omega_{480}	\\
0 												& \frac{1}{2}{\Omega_{475}}^*(t) & \frac{1}{2}{\Omega_{480}}^*	& - \Delta_{\mathrm{Ryd}}
\end{pmatrix}.
\end{equation}
The Lindblad operator $L(\hat{\rho})$, which is defined in analogy to its three-level version \cite{Huber2011a}, accounts for the spontaneous decays. The respective decay rates, however, do not have a significant influence on the timescales of the experiment.
We extract the time dependence of $\Omega_{475}(t)$ from the temporal envelope of the experimental pulse shape.

The radiated electric field due to the oscillating atomic dipole moment is determined by the coherence of the respective transition $\rho_{31}$. For the whole Doppler ensemble, the resulting electric field amplitude is given by the superposition of the individual fields which yields
\begin{equation}
\label{eq:efield}
E_0\propto N\cdot \left\langle\rho_{31}\right\rangle_\mathrm{v},
\end{equation} 
where $N$ is the atomic number density and the angular brackets $\left\langle\,.\,\right\rangle_\mathrm{v}$ denote the average over the one-dimensional Maxwell-Boltzmann velocity distribution.
If phase-matching is fulfilled, the intensity can be written as
\begin{equation}
I\propto N^2\cdot \left|\left<\rho_{31}\right>_v\right|^2.
\end{equation} 
As a consequence, interference effects between electric fields from different velocity classes can occur.

\section{Dynamics}

In the absence of the $475\,\mathrm{nm}$ pulse, the system is essentially uncoupled and thus resides in the ground state \ket{1}. During the pulse, the excitation dynamics occurs directly between the ground and Rydberg state due to the off-resonant intermediate state \ket{2}. The weak coupling of the transition $\ket{4}\rightarrow \ket{3}$ causes no additional dynamics in the system. 

\begin{figure}
\centering
\includegraphics[width=.9\columnwidth]{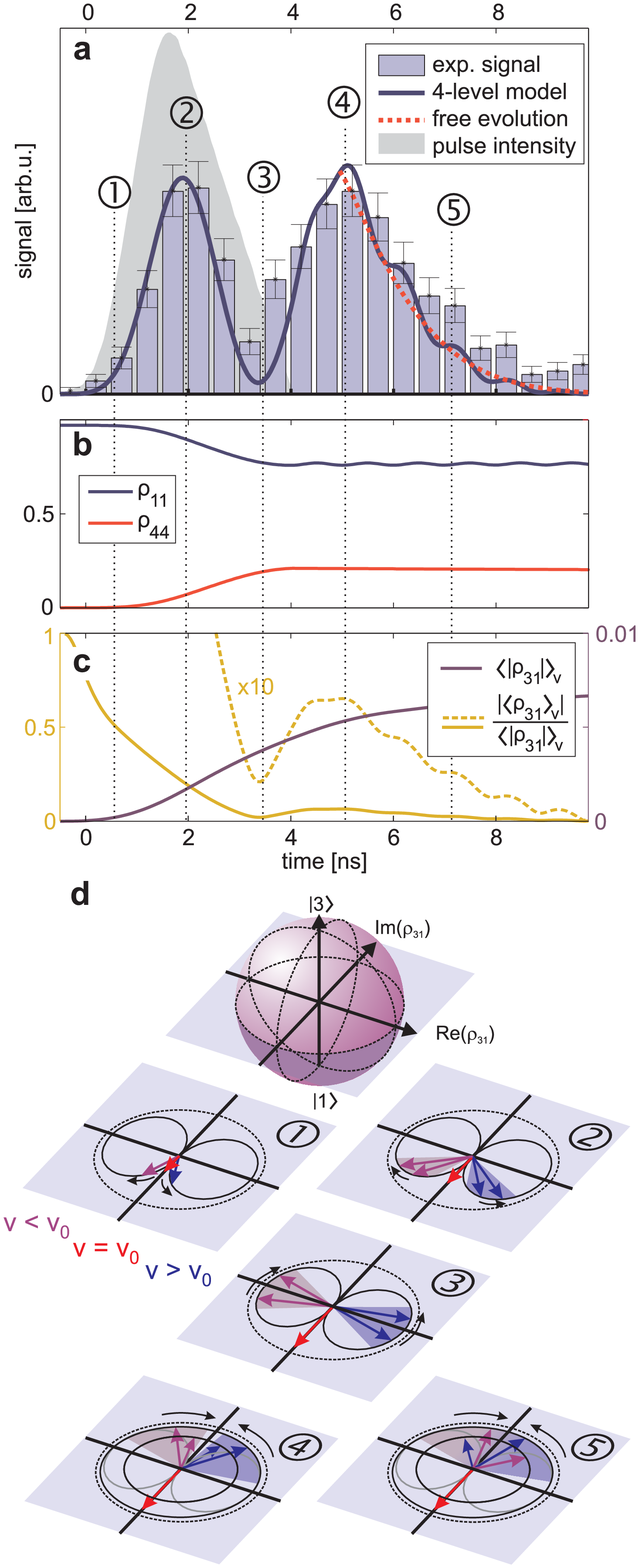}
\caption{(a) Typical four-wave mixing signal. The experimental signal (bars) is well described by a four-level model (blue line). After the pulse, the signal is governed by free evolution (red line) of the coherences in the thermal ensemble.
(b) Rydberg and ground state population for the velocity class $v_0$.
(c) Amplitude parameter $\left\langle\left|\rho_{31}\right|\right\rangle_v$ and interference parameter $\left|\left\langle\rho_{31}\right\rangle_v\right|/\left\langle\left|\rho_{31}\right|\right\rangle_v$. 
The interference parameter is also shown in 10-fold magnification (dashed line).
(d) Sketch of the coherences for different velocity classes in a Bloch sphere picture for selected points in time 
 (\textcircled{1}-\textcircled{5} indicated in (a)). Further explanations are found in the text.}
\label{fig:generic_oscillation}
\end{figure}

Within a certain range of Rabi frequencies we observe a double-peak structure in the temporal shape of the four-wave mixing signal (fig.~\ref{fig:generic_oscillation}a). In this case, the Rydberg state addressed is 30S at an atomic density  of $N = 0.4\,\mathrm{\upmu m}^{-3}$ \footnotemark[1]
, where no effects of Rydberg-Rydberg interaction are expected on the relevant timescale \cite{Baluktsian2013}. 
The error bars represent Poissonian uncertainties of the photon statistics. Fluctuations of the laser pulse intensity, which are on the order of 10\%, are not taken into account. 
As the individual Rabi frequencies on the two-photon transition to the Rydberg state are not completely negligible in comparison to the detuning to the intermediate state \ket{2}, the maximum contrast of the double-peak structure is not observed exactly at the two-photon resonance to the Rydberg state. By performing frequency scans of both upper lasers, the maximum contrast has been determined to be at $\Delta_\mathrm{Ryd}/2\pi = 200\,\mathrm{MHz}$ blue detuned to the Rydberg state (of atoms at rest).

The signal is well described by the four-level model, 
where the overall amplitude as been fitted to the data. The parameters of the model are $\Omega_{795}/2\pi = 335\,\mathrm{MHz}$, $\Omega_{475,\mathrm{max}}/2\pi = 375\,\mathrm{MHz}$ and $\Omega_{480}/2\pi = 5\,\mathrm{MHz}$, where $\Omega_{475,\mathrm{max}}$ denotes the peak pulse Rabi frequency. 
The Rabi frequencies agree with intensity and pulse energy measurements within the experimental error.
The effective two-photon Rabi frequency on the transition $\ket{1}\rightarrow \ket{4}$, in turn, is approximately given by $\Omega_\mathrm{eff}/2\pi = (\Omega_{795}\Omega_{475,\mathrm{max}}/2\Delta_{795})/2\pi = 63\,\mathrm{MHz}$.

\begin{figure}
\centering
\includegraphics[width=.9\columnwidth]{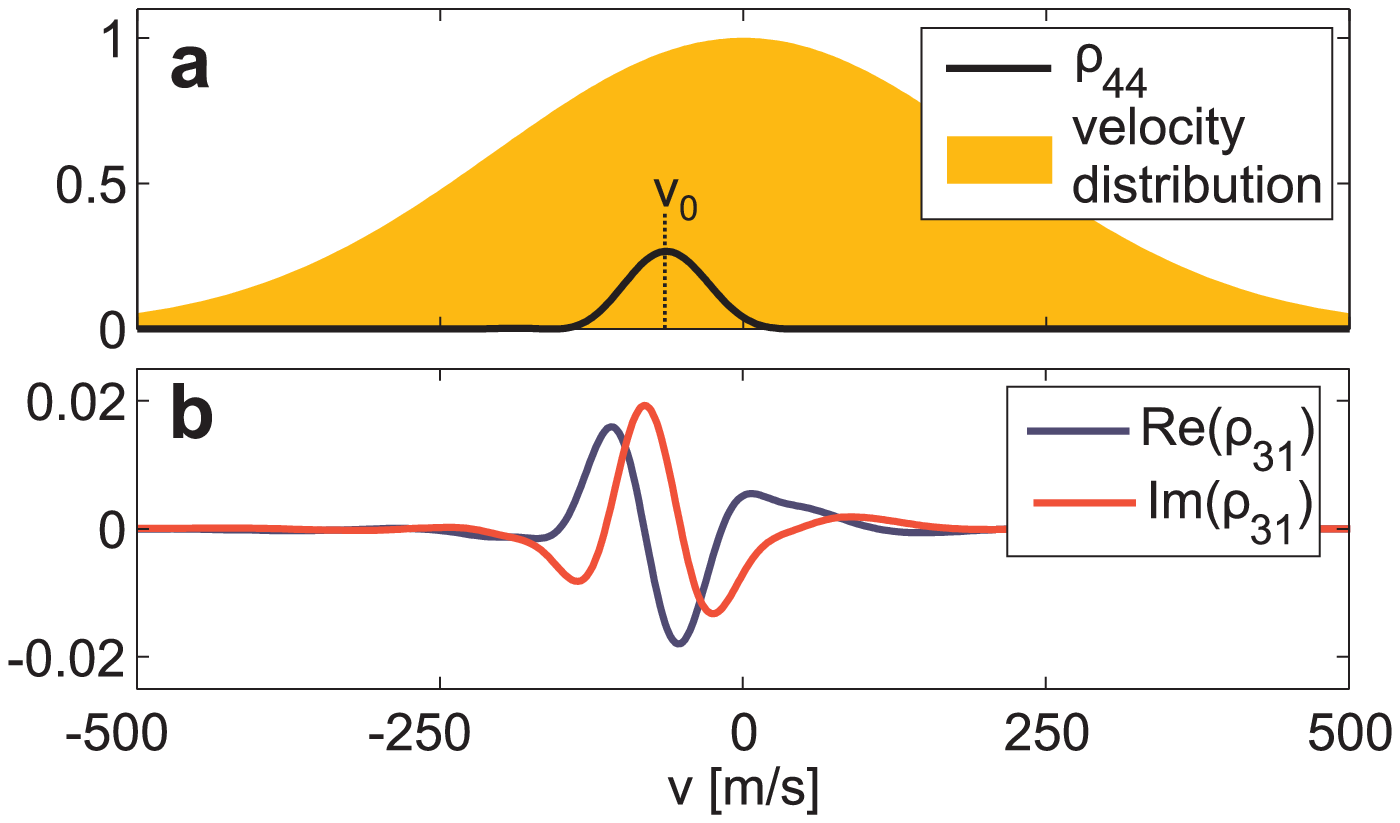}
\caption{
Selected atomic quantities for the different velocity classes in the thermal ensemble at $t_0 = 5\,\mathrm{ns}$ (\textcircled{4} indicated in fig.~\ref{fig:generic_oscillation}a). (a) Rydberg population and Gaussian velocity distribution. (b) Coherence on the radiating transition. 
}
\label{fig:generic_oscillation_vel}
\end{figure}

No Rabi oscillations are observed in the signal during the short time of the pulse in agreement with the small excitation Rabi frequency $\Omega_\mathrm{eff}$. The respective atomic populations extracted from the model (fig.~\ref{fig:generic_oscillation}b) for the velocity class of maximum Rydberg population $v_0$ 
show that the Rydberg population does not exceed $\sim 0.25$ corresponding to $1/8$ Rabi cycle ($\pi/4$ pulse). The second peak in the signal, therefore, cannot be attributed to Rabi oscillations but has to originate from the interplay of different velocity classes in the thermal atomic ensemble. 
Note that in this case $v_0\neq 0$ due to the finite $\Delta_\mathrm{Ryd}$ (fig.~\ref{fig:generic_oscillation_vel}a). 

In order to describe this behavior, we define two quantities: an amplitude parameter $\left\langle\left|\rho_{31}\right|\right\rangle_v$ that characterizes the magnitude of the radiated electric field of individual atoms in the Doppler ensemble and an interference parameter
\begin{equation}
	\frac{\left|\,\left\langle\rho_{31}\right\rangle_v\right|}{\left\langle\left|\rho_{31}\right|\right\rangle_v}
\end{equation} 
that accounts for the relative phase between the fields of different velocity classes. The interference parameter is normalized to one for fully constructive interference.
These two parameters can be extracted from the model and are shown in fig.~\ref{fig:generic_oscillation}c. 
At early times, the rise of the amplitude parameter is predominantly determined by the excitation dynamics to the Rydberg state due to the pulse. After the pulse ($t \approx 4\,\mathrm{ns}$), the curve flattens off and increases only slowly as additional populations from the Rydberg state is pumped down to state \ket{3} at the low coupling rate of the $480\,\mathrm{nm}$ laser. 
The interference parameter, on the other hand, is initially one as the time evolution of all velocity classes is still in phase for short times due to the Fourier uncertainty principle. The subsequent decline is caused by dephasing of different velocity classes since the time evolution is significantly determined by the Doppler detuning. The interference parameter reaches its smallest value at the same time as the signal minimum and exhibits a subsequent revival that is the cause of the second signal peak. 
The signal shape, hence, results from the combination of two effects: the increase of the individual coherences and simultaneous de- and re-phasing between different velocity classes.  

An intuitive illustration of the time-evolution can be given with the help of the Bloch sphere, where only the two levels of the radiating transition $\ket{3}\rightarrow\ket{1}$ are considered. 
Note, however, that the length of the corresponding Bloch vector is not constant in this case as the two levels do not form a closed system. 
In this picture, the coherences that describe the radiated electric field (eq.~\ref{eq:efield}) are the projection of the state vector into the equatorial plane.
The situation is sketched in the images in fig.~\ref{fig:generic_oscillation}d for characteristic points in time \textcircled{1}-\textcircled{5}, which are also indicated in fig.~\ref{fig:generic_oscillation}a. In order to maintain the symmetry of the illustration, the system is considered in the reference frame of the velocity class $v_0$.
As soon as the system is coupled by the laser pulse, the coherence of the velocity class $v=v_0$ (red arrow) is  increasing along the symmetry axis of the plane while the coherences of the faster and slower velocity classes (blue and violet arrows, respectively) move around ellipse-like trajectories right and left of the symmetry axis. The trajectories, again, are given by the projection of the respective Bloch vector trajectories onto the equatorial plane. (The fact that the aspect ratio of the trajectory is different for each velocity class has been neglected in the drawing for the sake of clarity. The dashed circles represent the boundary of the Bloch sphere.)
At the beginning of the pulse, the Bloch vectors of all velocity classes point towards the bottom of the sphere, such that the coherence is zero. As they start to rise, the coherences grow, leading to a signal increase (\textcircled{1}). Subsequently, the coherences dephase as they follow different trajectories determined by their respective atomic velocity. As a consequence, the signal increase slows down until a local maximum is reached (\textcircled{2}). Then, as the coherences evolve further, the signal decreases until a minimum is reached (\textcircled{3}). The minimum is predominantly caused by destructive interference of $v>v_0$ and $v<v_0$ velocity classes. Further evolution of the coherences causes the signal to increase again.
After the pulse, the levels are not coupled anymore and consequently all coherences perform circular motion along the equator while their magnitude, which for each velocity class is given by the prior time evolution due to the pulse, does not change anymore. This leads to a re-phasing on the opposite ``side'' (\textcircled{4}), which is the origin of the second signal peak. As the angular frequency depends on the velocity class, though, the subsequent dephasing (\textcircled{5}) finally causes the signal to decrease down to zero.

The exact relationship between the atomic dynamics and the signal shape after the pulse can be understood more easily in Fourier space.
In the case of zero coupling, the coherence of each velocity class evolves freely with the phase $\propto \exp(-i\cdot k_{780} v\cdot t)$, where $k_{780}$ in the phase-matched case is given by the wave vector of the signal light. Hence, the Doppler-averaged coherence can be written as
\begin{align}
	\rho_{31}(t) \propto \int dv  	&\;\overbrace{\exp\left(\scriptstyle{-\frac{mv^2}{2k_B T}}\right)\cdot\rho_{31}(t_0,v)   }^{A(v)}\nonumber\\
																	&\cdot \exp\left[-i\cdot k_{780} v\cdot (t-t_0)\right]. 
\end{align} 
This is essentially the Fourier transform of $A(v)$, which is given by the distribution of the coherences in velocity space at a certain time $t_0$ after the pulse. The coherences $\rho_{31}(t_0,v)$ themselves are determined by the prior atomic dynamics due to the pulse. The situation at time \textcircled{4} after the pulse is shown in fig.~\ref{fig:generic_oscillation_vel}. 
The excitation bandwidth determines the width of Rydberg-excited velocity classes $\Delta v$ (fig.~\ref{fig:generic_oscillation_vel}a) via the wave vector of the two-photon transition: $k_{\mathrm{eff}} \Delta v = (k_{795}+k_{475}) \Delta v$.
Due to the low excitation bandwidth, only a small window of velocity classes is excited to the Rydberg state out of the whole velocity distribution, 
which translates via the weak coupling of the $480\,\mathrm{nm}$ laser to a window in the velocity distribution of the coherences $\rho_{31}(t_0,v)$ (fig.~\ref{fig:generic_oscillation_vel}b). 
The signal ($\propto \left|\rho_{31}(t)\right|^2$) calculated from the Fourier transform of this distribution is plotted in fig.~\ref{fig:generic_oscillation}a and describes the decline of the signal in good agreement.
In general this description is valid as long as the additional population pumped down from the Rydberg state by the $480\,\mathrm{nm}$ laser after the pulse is small during the respective time interval.
In other words, the time evolution of the signal after the pulse is determined by the (phase and amplitude) aperture in momentum space $A(v)$ of the radiating atomic dipoles that has been imprinted by the excitation lasers. In our case, the aperture is quite smooth due to the low excitation Rabi frequency which, in turn, via Fourier transform leads to a smooth time-envelope of the signal that does not exhibit any additional revivals at later times.
In conclusion, the second signal peak originates from a re-occurrence of constructive interference between the coherences of different velocity classes. In that sense the second peak can be referred to as motion-induced revival. 

\begin{figure}
\centering
\includegraphics[width=\columnwidth]{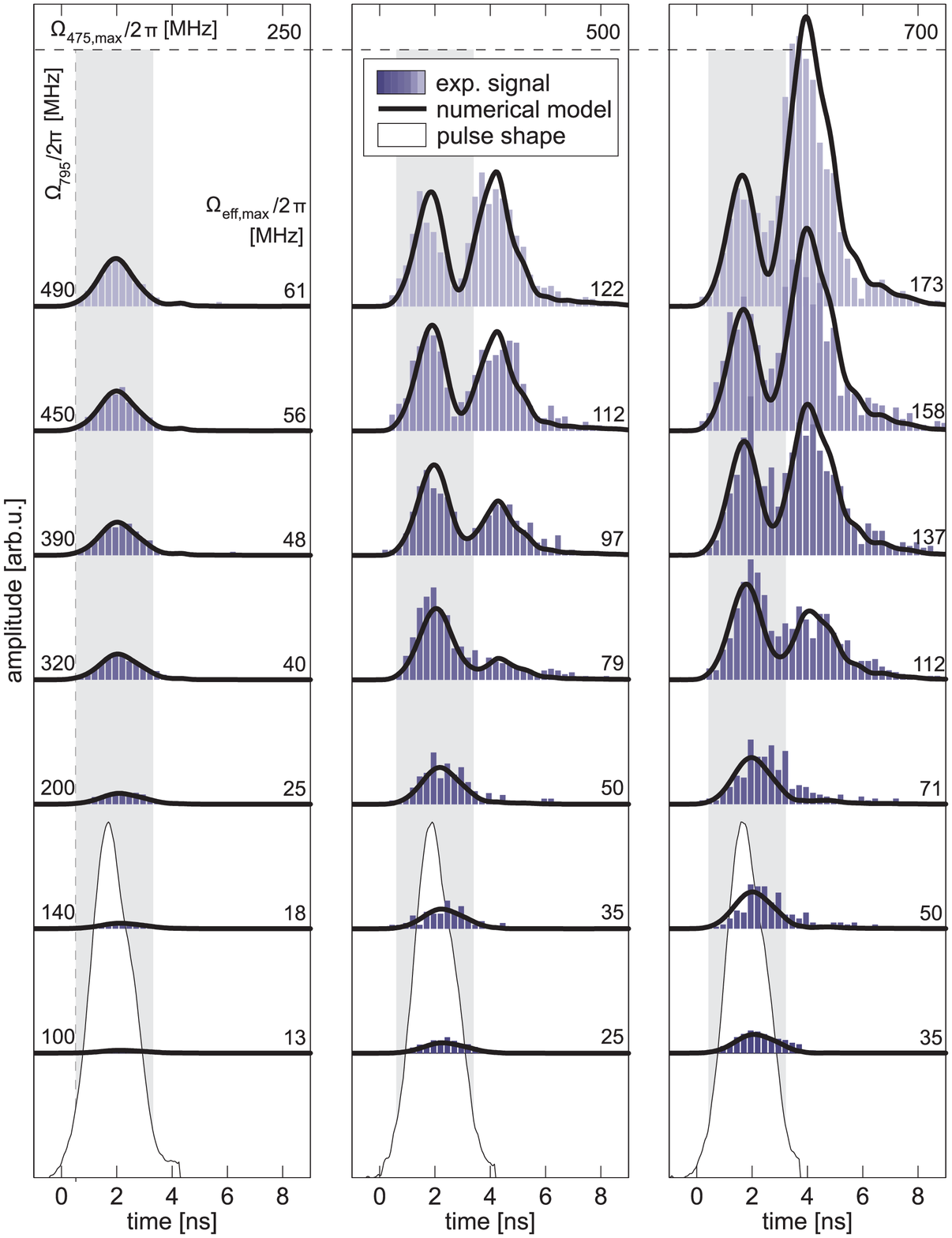}
\caption{Four-wave mixing signals for different excitation Rabi frequencies. The pulse Rabi frequency $\Omega_{475,\mathrm{max}}$ is varied along the columns and $\Omega_{795}$ along the rows. The envelope of the pulse intensity is depicted by the white-shaded curve at the bottom of each column together with a dark-shaded area to indicate the time interval of the pulse. 
\label{fig:Oscillations_bandwidth}}
\end{figure}

We have investigated the four-wave mixing signal for different excitation Rabi frequencies. 
Here, the Rydberg state addressed is 25S at an atomic density of $N = 1.9\,\mathrm{\upmu m}^{-3}$ \footnotemark[1], where no Rydberg interaction effects are expected.
The results are shown in fig.~\ref{fig:Oscillations_bandwidth} for different $\Omega_{795}$ (rows) and pulse Rabi frequencies $\Omega_{475,\mathrm{max}}$ (columns). The experimental data is well described by the four-level model over the whole range. The curves of the model are the result of a simultaneous fit to all experimental traces, where the only two fit parameters are the overall amplitude and a single parameter for the excitation Rabi frequencies $\Omega_{795}$ and $\Omega_{475,\mathrm{max}}$, which are related by a fixed ratio. This ratio is inferred from the relative laser intensities and pulse energies, taking into account the dipole matrix elements for the two different transitions. The temporal dependence of $\Omega_{475}(t)$ is entirely defined by the experimentally measured pulse shape.
Note that, as the dynamics is largely determined by the effective two-photon Rabi frequency $\Omega_\mathrm{eff} = \Omega_{795}\Omega_{475,\mathrm{max}}/2\Delta_{795}$, the temporal shape of the signal looks similar for different $\Omega_{795}$ and $\Omega_{475,\mathrm{max}}$ but same $\Omega_\mathrm{eff}$.

For small Rabi frequencies, only a single signal peak is observed. With increasing $\Omega_\mathrm{eff}$, the signal revival starts to occur and grows, finally exceeding the first peak. 
The height of the second peak is very sensitive to the effective Rabi frequency as it originates from interference as discussed above.
The relative height of the two peaks can thus be used as a convenient experimental tool to gauge Rabi frequencies.

The shape of the signal is determined mainly by the ratio between the excitation Rabi frequency $\Omega_\mathrm{eff}$ and the Fourier width of the pulse $\Delta\omega$.

In the case of $\Omega_\mathrm{eff} \ll \Delta\omega$, the excitation bandwidth is given by $\Delta\omega$ and the phase evolution of the coherences $\rho_{31}(t,v)$ is dominated by the respective Doppler detuning. 
Via Fourier transform, this translates to a single-peak envelope in the time-domain. In this regime of linear coupling, the coherence increases linearly with $\Omega_\mathrm{eff}$, i.e.\ the height of the signal peak is proportional to $\Omega_\mathrm{eff}^2$.

As the $\Omega_\mathrm{eff}$ approaches the same order of magnitude as $\Delta\omega$, the excitation bandwidth and, in turn, the time evolution of the coherences is determined by both quantities. The non-linear coupling favors the excitation of $v\neq v_0$ velocity classes slightly more than in the linear regime, which leads to the emergence of the signal revival as discussed above. 

An even higher Rabi frequency leads to a higher population of excited atoms. The increase of the second peak with $\Omega_\mathrm{eff}$, however, is predominantly caused by an enhancement of the constructive interference due to higher velocity classes participating in the process.
The height of the first peak does not increase significantly anymore at higher Rabi frequencies, as there are two effects that cancel each other. Specifically, the signal increases faster while the signal minimum shifts to earlier times due to higher velocity classes that are contributing.


In summary, the double peak structure in the four-wave mixing signal is clear evidence for coherent evolution of the individual velocity classes that are present in the hot gas of atoms.

Additionally, we have verified that the signal strength shows linear dependence on the $480\,\mathrm{nm}$ laser intensity ($\propto \Omega_{480}^2$) while the temporal shape remains unaffected.

\section{Density dependence}

Furthermore, we have investigated the signal strength of the four-wave mixing signal for different atomic densities. To avoid effects of Rydberg interaction, the Rydberg state 22S is chosen here. 
In this case, the pulsed laser is blue detuned by $360\;\mathrm{MHz}$ with respect to the two-photon resonance while the $480\,\mathrm{nm}$ laser is still on resonance. 

\begin{figure}
\centering
\includegraphics[width=\columnwidth]{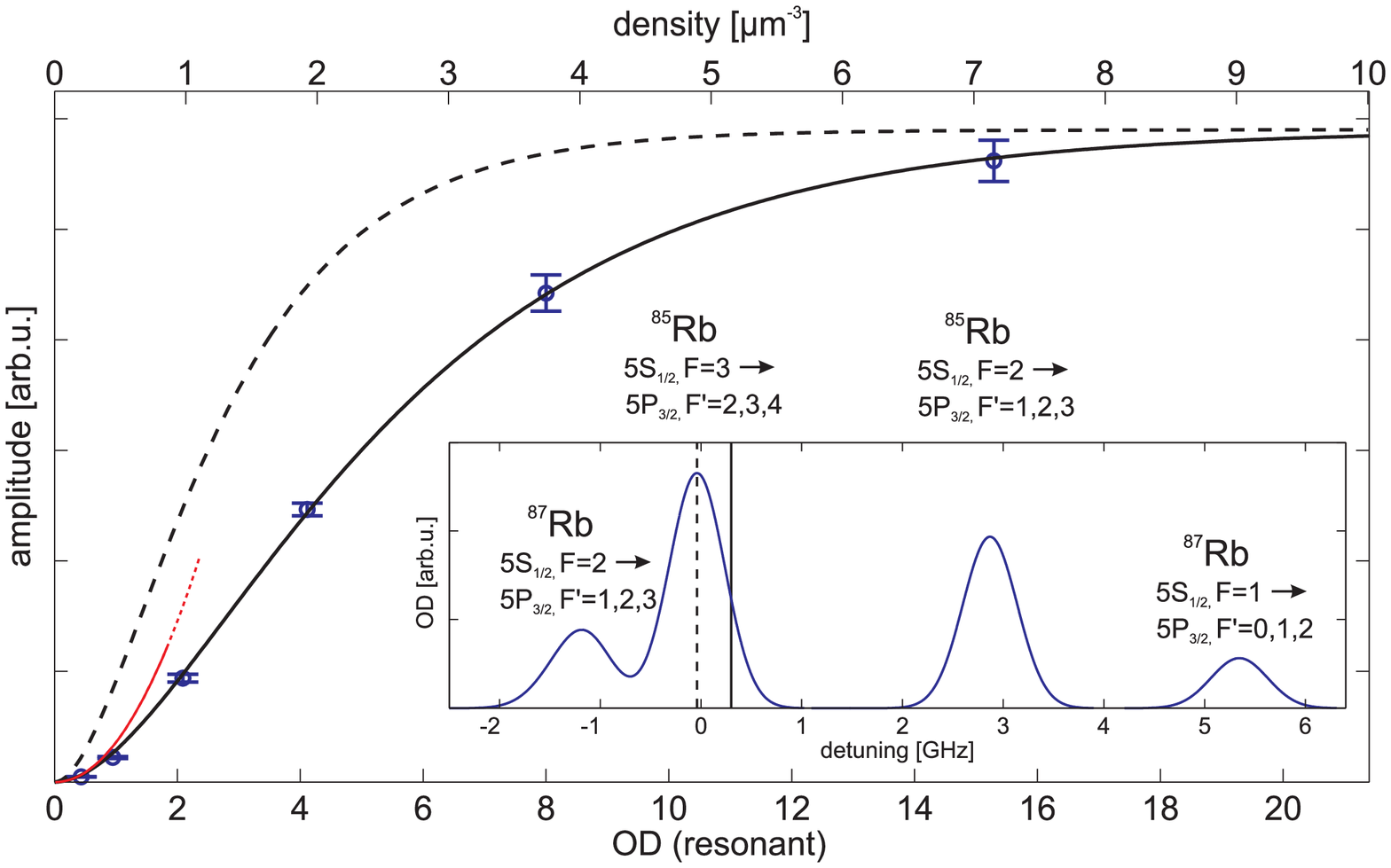}
\caption{Density dependence of the integrated four-wave mixing signal. The blue circles show the experimental data for different optical densities. The error bars are Poissonian standard deviations. 
The black lines represent the re-absorption model, where the optical density corresponds to a detuning of $360\;\mathrm{MHz}$ (black solid line) and to resonant light (black dashed line). The overall amplitude factor has been chosen to match the experimental saturation value. 
The red curve shows the model without re-absorption in the detuned case, which scales quadratically with the atomic number density.
$O\!D$ on the x-axis refers to the optical density for light resonant to the center of the Doppler profile of the $^{85}\mathrm{Rb}\; 5\mathrm{S}_{1/2}, F = 3\rightarrow 5\mathrm{P}_{3/2}, F'$ transition. The density axis refers to the density of $^{85}\mathrm{Rb}$ atoms in the $F = 3$ ground state. 
The inset depicts the optical density for the Rb D2 line. The dashed and solid vertical lines indicate the detunings, which correspond to the resonant and detuned models, respectively, illustrated in the main graph.
\label{fig:reabsorption}}
\end{figure}

Fig.~\ref{fig:reabsorption} shows the time-integrated four-wave mixing signal as a function of atomic and optical density. 
The signal saturates at high densities due to re-absorption of signal light in the atomic medium. For even higher densities ($N > 10\,\mathrm{\upmu m^{-3}}$), we find a decline in the signal, which is likely caused by superradiant effects of the Rydberg state but requires further investigation.

We describe the density dependence of the signal by means of a simple re-absorption model. Due to the off-resonant intermediate state \ket{2}, absorption effects on the excitation lasers can be neglected while they are passing through the atomic medium. The four-wave mixing signal itself, however, experiences absorption, which leads to the saturation behavior for high densities. For a medium with density $N$, optical density $O\!D$ and spatial extent $d$, a light field is created at each position $x$ in the medium along the direction of light propagation. This field is subject to absorption on its way through the remaining medium $d-x$. For sufficiently weak signals (Rabi cycle phase $\ll \pi$), the absorption can be described by Lambert-Beer's law. In case of phase-matching, the resulting electric field from all atoms is given by constructive superposition of the individual fields and yields
\begin{equation}
E_0\propto \int_0^d N \cdot\mathrm{exp}\left[-\frac{O\! D}{2d}(d-x)\right]\,\mathrm{d}x
\end{equation}
with the corresponding intensity 
\begin{equation}
\label{eq:reabsorption_intensity}
I\propto \left(\frac{N}{O\! D}\right)^2\left[1-\mathrm{exp}\left(-\frac{O\! D}{2}\right)\right]^2.
\end{equation}
The intensity curve follows a quadratic onset and subsequent saturation. 
The shape of the curve 
only depends on the absolute value of the optical density.
In general, the optical density is proportional to the atomic number density but also to the frequency of the respective light field (fig.~\ref{fig:reabsorption}, inset). Hence, for a known atomic density, the saturation behavior indicates the detuning of the four-wave mixing light. It can be seen from fig.~\ref{fig:reabsorption} that the experimental data agrees with the optical density of light detuned by $360\,\mathrm{MHz}$, which is consistent with the detuning of the laser pulse and thus with the conservation of photonic energy in the four-wave mixing process. The absorption curve for resonant light clearly does not fit the data. Note that in both cases, the amplitude has been fitted to the saturation value of the experimental signal. When comparing absolute amplitudes, both curves follow the same onset but the curve for resonant light saturates earlier (at $(N/O\! D)^2$).

Furthermore, the data exhibits a quadratic onset that is reproduced by the model in very good agreement. The quadratic onset is a direct consequence of phase-matching, i.e.\ momentum conservation for the light fields.
In summary, we find consistency of the four-wave mixing signal with momentum and energy conservation for the participating light fields from the position of the slope and the quadratic signal onset.

\vspace{3cm}
\section{Conclusion}

In conclusion, we have observed time-resolved signals from a pulsed four-wave mixing process via a Rydberg state on the nanosecond timescale in thermal vapor of rubidium. 
While previous experiments have demonstrated that coherent Rydberg dynamics can be achieved in the frozen gas regime,  where all atoms evolve equally \cite{Huber2011a}, we have investigated the regime, where the atoms evolve differently depending on the respective velocity as the Doppler effect has a significant influence on the temporal evolution. 
We have observed a revival of the four-wave mixing signal that we can attribute to re-phasing of different radiating atomic velocity classes in the thermal ensemble. The signal revival is thus evidence for coherent Rydberg evolution  beyond the frozen gas regime in an Alkali gas above room temperature. 
Over a large range of Rabi frequencies, we find good agreement with a four-level model.

Finally we have investigated the density dependence of the four-wave mixing signal. By comparing the shape of the corresponding curve to a re-absorption model we have been able to draw conclusions about the phase-matching and the detuning of the four-wave mixing light, which is consistent with energy and momentum conservation of the light fields involved.





\begin{acknowledgments}
The work is supported by the ERC under Contract No. 267100, BMBF within Q.com-Q (Project No. 16KIS0129) and the EU project MALICIA. B.H. and A.K. contributed equally to this work.
\end{acknowledgments}


%

\end{document}